# Coherent signal detection in the statistical polarization regime enables high-resolution nanoscale NMR spectroscopy

Nick R. von Grafenstein[1,2]†, Karl D. Briegel[1,2]†, Pol Alsina-Bolívar[3,4], Jorge Casanova[3,4]

Dominik B. Bucher[1,2*]

[1] Technical University of Munich, TUM School of Natural Sciences, Chemistry Department, Lichtenbergstraße 4, 85748 Garching, Germany
[2] Munich Center for Quantum Science and Technology (MCQST), Schellingstr. 4, 80799 Munich, Germany
[3] Department of Physical Chemistry, University of the Basque Country UPV/EHU, Apartado 644, 48080 Bilbao, Spain
[4] EHU Quantum Center, University of the Basque Country UPV/EHU, Bilbao, Spain

†These authors contributed equally to this work
*Corresponding author: dominik.bucher@tum.de

## Abstract

**Nitrogen-vacancy (NV) centers in diamond have emerged as quantum sensors capable of detecting nuclear magnetic resonance (NMR) signals at unprecedented length scales, ranging from picoliter sample volumes down to single spins at the diamond surface. While high-resolution (few-hertz) coherent NV-NMR spectroscopy has been demonstrated at the micrometer scale, achieving comparable resolution at the nanometer scale has remained elusive. In this nanoscale regime, sensing relies on detecting stochastic spin noise from statistical polarization, where molecular diffusion in liquid samples severely broadens the spectrum and degrades resolution. Here, we demonstrate that coherent detection of signals from a hyperpolarized nanoscale sample overcomes this limitation, enabling single-digit hertz spectral resolution and the capacity to resolve scalar couplings. Finally, building on these results, we present a comparative analysis of the sensitivity of coherent versus spin-noise detection, identifying the regimes in which each performs best.**

## Main text

Nuclear Magnetic Resonance (NMR) spectroscopy is at the forefront of analytical techniques in modern science, elucidating molecular structure and dynamics and providing insight into interactions at the atomic level. Its applicability to research fields as diverse as chemistry, medicine, and materials science, combined with its non-destructive nature, has made it an indispensable tool. However, traditional induction-based NMR spectroscopy requires substantial sample volumes, which precludes its use for nanoscale samples. In recent years, nitrogen-vacancy (NV) centers in diamonds[1,2] have emerged as quantum sensors capable of detecting NMR signals at unprecedentedly small scales[3–5]. NV-NMR captures the interaction between the spins of the sample and the NV-centers, transforming this interaction into an optical signal, encoded in the NV-center's fluorescence intensity[1,2,6,7]. This interaction exhibits a strong dependence on distance, meaning that the dominant detection volume of an NV center is determined by its depth below the diamond's surface[8–11]. Consequently, NV-NMR applications have historically fallen into one of the two categories:

1) Microscale NV-NMR spectroscopy relies on NV ensembles extending over several micrometer-thick layers on the surface of a diamond[12–18]. This places the dominant detection volume above the ensemble in the picoliter range for single-cell[19] or lab-on-a-chip studies[20] (Fig. 1a). At this scale, coherent detection is allowed since the spin polarization of the sample is typically governed by the alignment of the spins with a magnetic bias field $B_0$ following the Boltzmann distribution (i.e. thermal polarization, $P_{therm} \sim \frac{B_0}{k_B T}$) or by hyperpolarization techniques (e.g., Overhauser dynamic nuclear polarization (ODNP)[13] or parahydrogen based methods[14]). Using phase-sensitive NV detection protocols (such as the coherently averaged synchronized readout (CASR) protocol[12]), coherent NMR signals can be recorded with remarkable frequency resolution[12,21,22]. In this case, the signal linewidth is primarily limited by the coherence time of the sample spins.

2) Nanoscale NV-NMR spectroscopy techniques, in which either a single NV[9,11,23–26] or an ensemble of NVs[27–31] is placed a few nanometers from the diamond surface (Fig. 1b). These techniques are critical for analyzing extremely limited amounts of samples, such as single monolayers of material on a surface[24,28] or even single molecules and spins[25,32,33]. In this case, the NV center interacts with only a small number N of spins (few 1000) whose spin polarization is dominated by statistical fluctuations of the spin bath (i.e., statistical polarization, $P_S \sim \frac{1}{\sqrt{N}}$, Fig. 1c)[34–36]. These signals arise spontaneously, eliminating the need for active radiofrequency (RF) excitation, and can be readily detected using noise spectroscopy[6,37,38] techniques (i.e., spin-noise sensing) by NV centers positioned a few nanometers beneath the diamond surface. However, due to their stochastic nature, the lifetime of the detected signal is limited by the diffusion of sample nuclei through the dominant detection volume, limiting the achievable linewidths to several hundred hertz for highly viscous samples to megahertz linewidths for liquid samples[9,10,30,37,39,40]. One solution is to confine the liquid sample so that the spins cannot diffuse out of the detection volume, which has been proposed theoretically[41] and demonstrated experimentally[42] recently.

Another approach involves detecting coherent signals from a uniformly polarized (i.e. thermal or hyperpolarized) sample[43]. In this regime, the NMR signal becomes independent of diffusion due to the coherence of the sample spins leaving and entering the nanoscale detection volume, enabling high spectral resolution. Another key advantage is the enhanced sensitivity of NV schemes that detect coherent NMR signals (slope detection) compared to noise spectroscopy applied to statistical polarization (variance detection)[3,6]. Furthermore, the NMR signal strength of a uniformly polarized sample on the diamond does not change if the sensor-to-sample geometry remains constant (geometry factor)[8].

In this Letter, we present the experimental demonstration of coherent NMR signal detection from hyperpolarized liquid samples at the nanoscale, a regime where statistical polarization typically dominates. Using an ensemble of shallow NV centers, we achieve NMR signals with single-digit hertz linewidths and resolve scalar couplings with high sensitivity. Furthermore, we complement these experimental findings with a comparative study between coherent detection and spin-noise sensing, demonstrating that the polarization level achieved in our experiment ($\sim 5 \cdot 10^{-5}$) is already sufficient for coherent detection to surpass the sensitivity limits of spin-noise sensing.

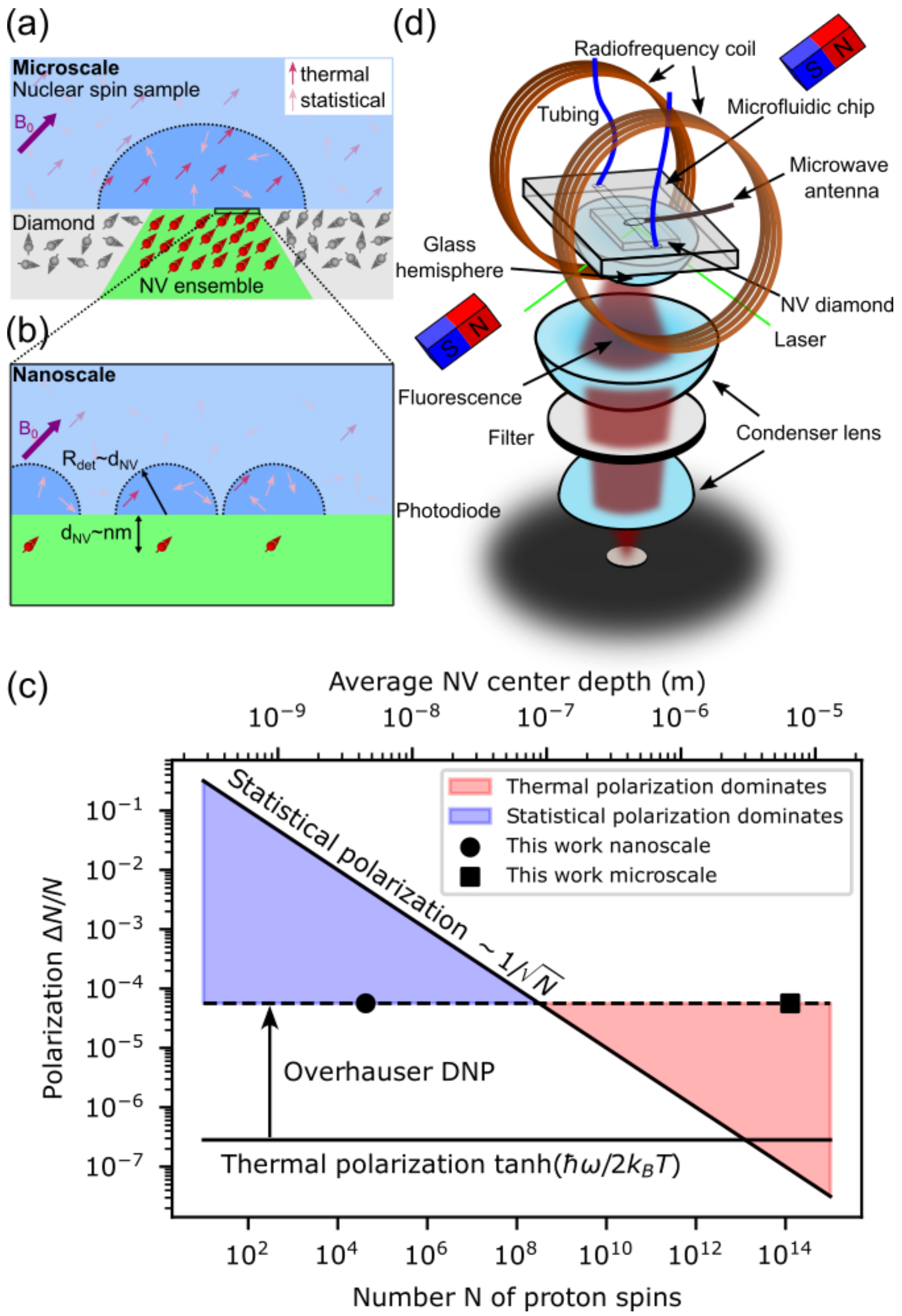

**Figure 1. Different regimes of NV-NMR spectroscopy.** Illustration of a) micro- and b) nanoscale NV-NMR experiments. The NV detection radius $R_{det}$ is proportional to the average NV center depth $d_{NV}$. c) Comparison of statistical- and thermal proton spin polarization at 84 mT and 300 K as a function of number of spins $N$. The Overhauser hyperpolarization enhancement of water is ~ 200[13]. d) Schematic of the experimental setup used for high-resolution nanoscale NV-NMR spectroscopy.

**Simulations.** At the nanoscale, the detection of statistical polarization within the small sensing volume of each NV center amplify the effects of molecular diffusion, giving rise to two main effects: (1) Nuclear spins with different initial states continuously enter and leave the sensing region. This continuous exchange of spins leads to a loss of coherence in the detected NMR signal and, consequently, to a reduction in spectral resolution. (2) Due to the spatial dependence of the NV–nuclear spin interaction (see below), a single spin diffusing through the effective sensing volume may experience a change in the sign of the interaction. We refer to these events as *phase jumps*. It is straightforward to see that coherent detection at the nanoscale, enabled for example by hyperpolarization, eliminates the first diffusion-related effect (1)[43]. However, phase jumps still occur even when all nuclear spins share the same initial state, i.e., in a homogeneously polarised sample. Moreover, at the nanoscale, these events typically occur on timescales much shorter than the nuclear Larmor period. Then, we firstly investigate whether phase jumps could compromise the coherent detection of intrinsic nuclear properties, such as chemical shifts and J-couplings, in a diffusing, homogeneously polarised nanoscale sample. For completeness, we perform the same analysis in the microscale regime. As shown below, we find that phase jumps do not degrade the detected signal, even in the limiting case where only a single nuclear spin occupies each sensing volume.

For the simulations, we employ a magnetic dipole-dipole model to describe the interaction between the NV center spin and the nuclear spins of the sample[8,12]**.** The interaction strength *S(t)* between the sample spin and the NV-center's spin is approximated by the projection of the sample spin's magnetic field onto the NV center's quantization axis, given by

$$S(t) = \frac{\mu_0}{4\pi}\left[\frac{3\hat{r}\left(\hat{r}\cdot{}^{1}\vec{H}(t)\right)-{}^{1}\vec{H}(t)}{r^3}\right]\cdot\overrightarrow{NV} \qquad \text{(Eq. 1)}$$

where $\mu_0$ is the magnetic permeability, $\hat{r}$ normalized distance vector between NV and nucleus, r scalar distance, ${}^{1}\vec{H}$ and $\overrightarrow{NV}$ denote the orientation of the sample spin and the NV axis, respectively. We note that our simulation focuses on the qualitative behavior and time evolution, rather than the absolute values of the physical parameters (see SM 4).

The dipolar interaction divides the detection volume into different regions where the sign of the interaction is opposite, causing the NMR signal phase to invert accordingly (Fig. 2a). To highlight the distinct impact of phase jumps in the micro- and nanoscale regimes, we simulate the interaction between a single NV center and a single diffusing nuclear spin in both scenarios. The sample region is created with a computer-aided design software and consists of a hemisphere centered above the NV center, with a radius equal to ten times the NV depth. As depicted in the trajectories in (Fig. 2a), the simulations are performed by initializing a single nuclear spin at a random position within the sample volume and assigning it an initial spin orientation in the equatorial plane. The spin then undergoes diffusion through the detection hemisphere, during which phase jumps may occur (if the spin crossed from, e.g. the red to the blue region). Specifically, at each time step, after calculating the interaction strength between the sample spin and the NV center, the nuclear spin position is updated by adding a random displacement vector, drawn from a Gaussian distribution parameterized to describe its diffusion. Additionally, the sample spin orientation is rotated around the $B_0$ field axis according to its Larmor frequency. Eventually, the spin may diffuse beyond the dominant detection volume such that the NMR signal becomes negligible. At this point, a new and equally polarized sample spin enters the detection volume in order to keep the nuclear spin concentration.

Note that in typical microscale NV-NMR experiments, since detection volume is large relative to the diffusion length (Fig. 2a, left), the sample spin's precession frequency is significantly faster than its diffusion through the detection volume, effectively keeping its position nearly stationary during each precession cycle (no phase jumps occur). That is, a single sample spin rarely crosses these regions, which appears as distinct phase jumps in the NMR signal (Fig. 2b, left). Moreover, the signal amplitude is modulated by the $1/r^3$ distance dependence on a timescale slower than the spin precession. Summing many random realizations of this process, thereby simulating a sensing volume populated by nuclear spins, and subsequently applying the CASR detection scheme[12] which transforms the oscillating NMR signal at the NV center into an optical signal oscillating at an aliased frequency, preserves both the slow drifts and the phase jumps while revealing the intrinsic oscillation frequencies (e.g. chemical shifts and J-couplings) (Fig. 2c, left).

In typical nanoscale experiments diffusion through the sample volume occurs significantly faster than sample spin precession (Fig. 2a, right). Here, the diffusion affects the amplitude and phase of the NMR signals within a single precession cycle – on a much faster timescale compared to the microscale experiment (Fig. 2b, right). Thus, for a single realization (single spin diffusing through the detection volume), no coherent information can be extracted from the dipolar interaction, and thus from the NV after applying the CASR protocol. However, after averaging over many random realizations of such events, and applying the CASR pulse sequence to the NV, only the coherent information is left on the state of the NV, while the fast “noise” from the phase jumps is averaged out due to its randomness (Fig. 2c, right).

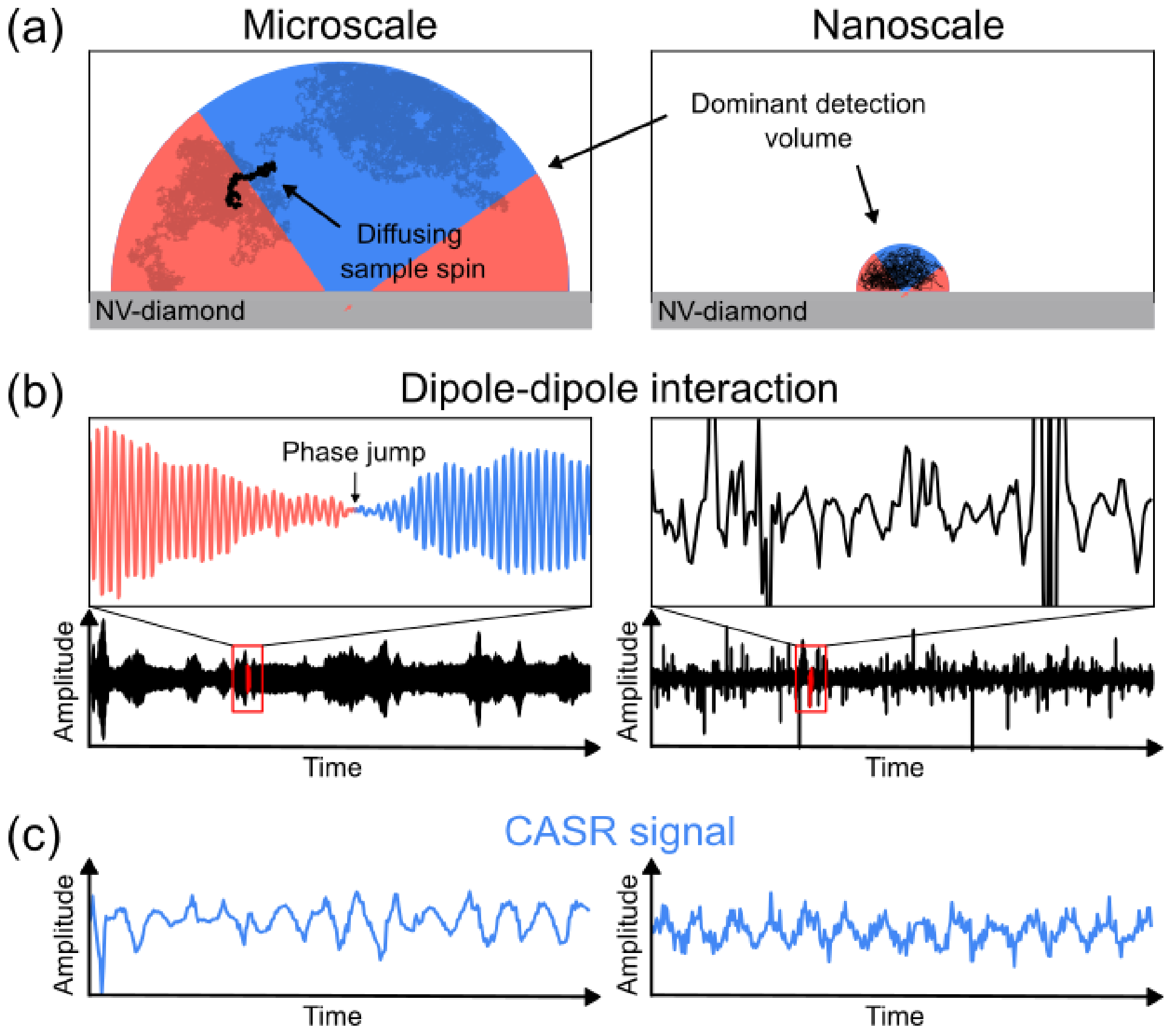

**Figure 2. Simulations of coherent signal detection within micro- and nanoscale detection volumes**. (a) Schematic representation of a sample spin diffusing through the NV center’s detection volume. The left and right panels illustrate the micrometer- and nanometer-scale NV-NMR regimes, respectively. The right nanoscale panel is scaled up for better visualization. The dipolar interaction divides the detection volume into regions with opposite interaction signs (red and blue), which impacts the NMR signal. Black lines indicate one diffusive trajectory of a spin within such detection volume. (b) Simulated dipole-dipole interaction strength over time, showing amplitude and phase variations due to sample spin diffusion. On the micrometer scale, where diffusion through the sample volume is relatively slow compared to the precession frequency, the diffusion gradually modulates the NMR signal’s amplitude and phase. On the nanometer scale, diffusion occurs much faster than Larmor precession, causing rapid amplitude and phase modulation. The zoomed-in sections of the left and right panels span different time intervals to better illustrate each regime's behavior. (c) Signals detected by the NV center using the CASR pulse sequence after averaging over many random realizations. The amplitudes are scaled differently relative to each other and do not depict physical values.

Importantly, we find that even in the unfavorable case of a single spin, the diffusion between regions of positive and negative dipolar interaction is not prohibitive to signal acquisition when performing coherent detection of the NMR signal. From the perspective of a specific NV-center, the interaction strength and even the sign of an individual diffusing sample spin may fluctuate over time. While signals with opposing signs partially cancel each other out, the weighted sum from the different regions, determined by the geometry factor[8], still contributes to the net NMR signal, as long as the number of performed realizations is enough (number of spins within the detection volume, number of NVs or number of experimental averages). These simulations show that the timescale of diffusion across regions with different interaction signs, blue and red in Fig. 2a and 2b, affects individual spin interactions but does not compromise the NV center's ability to acquire a signal. Averaging either over time or through multiple sample spins or NVs, the phase and amplitude variations average out, minimizing differences in signal shape between the nano- and microscale regime.

**Experimental Methods.** The experimental setup is shown in Fig. 1d. The NV-NMR spectrometer is based on a synthetic diamond chip containing NV centers positioned ~ 4.5 nm (2.5 keV implant energy, particle fluence of $2 \times 10^{12}/cm^2$ ) below the diamond surface, with an estimated density of ~ 50 - 100 NVs/$\mu m^2$, where the dominant detection volume of most NVs is isolated and not overlapping[28]. As a reference for the microscale regime, we use a diamond with a ~ 13 μm NV doped layer[17]. The diamond chip is bonded to a microfluidic glass chip for sample delivery and control[20]. This assembly is mounted into a 3D-printed sample holder equipped with copper coils for RF signal delivery and placed into a highly homogeneous magnetic field (~ 84 mT) originating from a temperature-stabilized, custom-designed permanent magnet[18,44]. NV control fields are delivered from a microwave (MW) loop antenna, positioned above the microfluidic assembly[45]. We further increase the thermal polarization by ODNP (see SM 1)[13,46] using 4-Hydroxy-2,2,6,6-tetramethylpiperidine-1-oxyl (TEMPOL) as a hyperpolarization agent. We then excite the hyperpolarized NMR sample with a $\pi/2$-pulse, which initiates the precession of the nuclear sample spins around the applied magnetic field. The resulting NMR signal is detected by the NV centers using the CASR protocol.

**Experimental results.** We begin with a micrometer-scale NV diamond sensor, which allows for high signal-to-noise ratio (SNR) single-shot measurements for comparison and calibration of the experiment (see SM 1). Once the parameters (e.g., $\pi/2$-pulse duration) are determined, we replace the micrometer-scale NV sensor with a nanometer-scale NV sensor, while keeping the rest of the setup unchanged, resulting in a pre-calibrated experiment. We would like to emphasize that changing the NV depth scales the detection volume while maintaining a constant NMR signal amplitude at the NV center for a uniformly polarized sample under the same geometry[8,12]. In the nanoscale regime, large NMR signals due to statistical polarization dominate; however, due to the NV ensemble and NV detection scheme (CASR), non-coherent signals are averaged out. We observe an NMR signal with linewidths in the range of ~ 5 - 6 Hz (~ 1.5 ppm) (Fig. 3a and 3b) in our experiments, representing an improvement of around three orders of magnitude over non-viscous samples at the nanoscale compared to previous work[42]. Additional experiments prove that we indeed see the signal from our hyperpolarized water sample (see SM 2).

For the water sample (Fig. 3a), we achieve an SNR of ~ 35 after 720 seconds averaging time, which corresponds to an estimated proton spin number sensitivity of ~ 20 fmol $Hz^{-1/2}$, based on $3 \times 10^4$ individual dominant NV detection volumes (see SM 1). Considering the sensor area (i.e., the excitation laser spot size), the proton spin number sensitivity is ~ 870 fmol $Hz^{-1/2}$. In future applications, nanofluidics[47] will be needed to exploit the small detection volume of our sensors.

Based on these results, we demonstrate scalar coupling resolved NMR spectroscopy detecting the $J_{HP}$ in trimethylphosphate (TMP), well resolving the scalar coupling constant of ~ 10 Hz (Fig. 3b). For TMP we obtain an SNR of ~ 10 after 2160 seconds averaging time. The lower sensitivity is attributed to less efficient ODNP enhancement[13] and splitting of the resonance line. Finally, we compare the NMR spectra from the nanoscale regime with those from the microscale regime (see SM 3). In both cases, the linewidths are very similar, suggesting that the limiting factors are likely magnetic field inhomogeneities, temperature-induced magnetic field drifts, and relaxation enhancement due to the paramagnetic TEMPOL in the sample solution, rather than effects at the nanoscale.

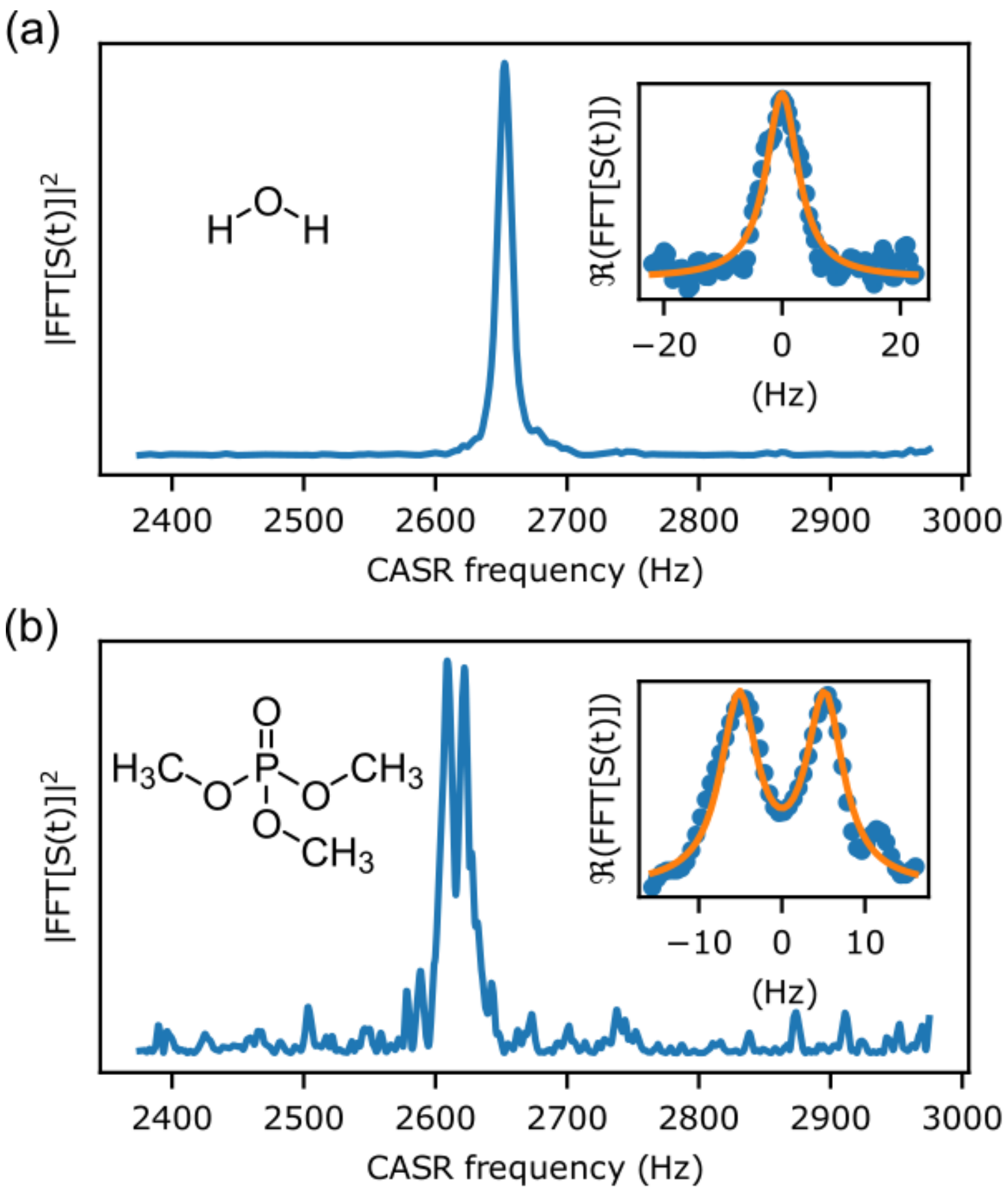

**Figure 3. Nanoscale NV-NMR spectra of liquid samples.** a) NV-NMR power spectrum of water (400 averages, 720 seconds total measurement time). *Inset:* NV-NMR spectrum (real part of Fourier transform) of water. Lorentzian fit (orange line) reveals a full width at half maximum (FWHM) linewidth of ~ 6 Hz. The frequency axis has been set to zero at the center of the resonance. b) Scalar coupling resolved NV-NMR power spectrum of trimethyl phosphate (TMP) (1200 averages, 2160 seconds total measurement time). *Inset*: NV-NMR spectrum (real part of Fourier transform) of TMP. Lorentzian fit (orange line) reveals an FWHM linewidth of ~ 5 Hz and scalar coupling constant of ~ 10 Hz. The frequency axis has been set to zero at the center of the resonance.

**SNR comparison of the CASR and Qdyne protocols.** The conventional approach for studying nanoscale samples with NV centers is to detect the statistical polarization, as it provides a much stronger signal than that obtained via thermal mean polarization (see Fig.1). As demonstrated experimentally in the previous section, hyperpolarization enables the coherent detection of the mean polarization even at the nanoscale. In terms of spectral resolution, this approach is inherently superior because molecular diffusion does not limit the coherent detection of a homogeneously polarized sample. An important question, however, is how its SNR compares with that of statistical polarization detection under identical experimental conditions, assuming that the sample is confined[42], to avoid diffusional line broadening. In this section, we numerically compare the SNR achieved with our approach, based on hyperpolarization and CASR, to that obtained with Qdyne[21,22,37], which typically detects statistical polarization. More specifically, our goal is to find for which values of the sample polarization and coherence time $T_2^*$ it is better (in terms of SNR) to use CASR or Qdyne (i.e. detect mean sample polarization or statistical polarization). To extend Qdyne to NV ensembles, we assume that the photons emitted by each NV center can be collected independently, e.g. on a camera.
The power spectrum SNR of CASR and Qdyne is (derivation in SM 6)

$$\mathrm{SNR_{CASR}} \propto \frac{p^2\, M_{CASR}\cdot K}{(\sigma_\xi)^2} f_s T_2^* \qquad \mathrm{SNR_{Qdyne}} \propto \frac{\sqrt{M_{qdyne}\cdot K}}{(\sigma_\xi)^2} f_s T_2^*, \qquad \text{(Eq. 2)}$$

where $p$ is the sample mean polarization (thermal in the case of no hyperpolarization), $M_{CASR}$ and $M_{Qdyne}$ are the number of experimental averages and $K$ is the number of NV sensors ($K \sim 3\cdot 10^4$ for our ensemble), $f_s$ is the sampling frequency and $T_2^*$ the dephasing time of the sample. $\sigma_\xi$ is the standard deviation associated to all the noise sources in the experiment. Note that $M_{CASR} \neq M_{Qdyne}$, because in CASR one must wait $T_1$ for the sample to thermalize before performing the next realization, whereas in Qdyne there is no such limitation. In addition, CASR needs to include the overhead time due to the Overhauser polarization time, $t_{ovh} = 0.3\ s$ in our experiment. Thus, $M_{CASR} = t_{tot}/(t_{ovh} + T_1)$ and $M_{Qdyne} = t_{tot}/T_2^*$. Then, (Eq. 2) is

$$\mathrm{SNR_{CASR}} \propto \frac{p^2\cdot K}{(\sigma_\xi)^2}\frac{t_{tot}}{t_{ovh}+T_1} T_2^* \qquad \mathrm{SNR_{Qdyne}} \propto \frac{\sqrt{K}}{(\sigma_\xi)^2}\sqrt{\frac{t_{tot}\,T_2^*}{2}}. \qquad \text{(Eq. 3)}$$

It is worth highlighting that the two detection schemes exhibit fundamentally different averaging behavior. Since CASR is a linear magnetometry technique, averaging is performed directly on the time-domain signal, yielding an SNR that scales as $\propto \sqrt{K\cdot\frac{t_{tot}}{t_{ovh}+T_1}T_2^*}$. When the power spectrum is then computed, the corresponding spectral SNR scales linearly with the number of averages, $\mathrm{SNR_{CASR}} \sim K\cdot\frac{t_{tot}}{t_{ovh}+T_1}T_2^*$. In contrast, Qdyne is based on statistical polarization, so the autocorrelation must be computed before averaging over repeated measurements. As a result, the averaged quantity already corresponds to the power spectrum, leading to the scaling $\propto \sqrt{K\frac{t_{tot}\,T_2^*}{2}}$. In Figure 4a and b we present the results of numerical simulations corresponding to the results for $K = 3\cdot 10^4$ and $t_{tot} = 720$ s. For completeness, in panels 4c and d we depict the results for single NV detection of the same sample (i.e., $K = 1$) and $t_{tot} = 39$ h. The rest of the parameters are extracted from the experimental results shown in Fig. 3: $t_{ovh} = 0.3\ s$, $T_1 = 1\ s$ and $\sigma_\xi \approx 6$ (see SI 6). Figure 4a shows a density plot of $\log_{10}(\mathrm{SNR_{CASR}})$, in terms of the sample mean polarization $p$ (horizontal axis) and to the dephasing time of the sample $T_2^*$ (vertical axis). The warm region corresponds to $\log_{10}(\mathrm{SNR_{CASR}}) < 0 \rightarrow \mathrm{SNR_{CASR}} < 1$, whereas the cold colour region corresponds to $\log_{10}(\mathrm{SNR_{CASR}}) > 0 \rightarrow \mathrm{SNR_{CASR}} > 1$. The black dotted line indicates the points where $\mathrm{SNR_{CASR}} = \mathrm{SNR_{Qdyne}}$, see (Eq. 3). As demonstrated by the experiment and the simulations, for our experimental parameters ($K = 3\cdot 10^4$, $\sigma_\xi \approx 6$, $t_{tot} = 12$ min) the achieved polarization $p \approx 5\cdot 10^{-5}$ is already sufficient for CASR to resolve with good SNR ($\sim 35$) (yellow star in the density plot). In addition, the SNR is higher than that obtained from the simulation of the Qdyne protocol, where statistical polarization is detected, since the yellow star lies at the right of the dotted line. Equation 3 predicts that, for a fixed value of $T_2^*$, the SNR scales as $\mathrm{SNR_{CASR}} \propto p^2$. The numerical simulations shown in (Fig. 4b) confirm this prediction: in a logarithmic representation, the SNR depends linearly on $p$ with the expected slope of 2. We believe that the loss of linearity for high values of $p$ is due to the breakdown of the small angle condition, required for the correct NV detection. Here, the dashed red line at high values of $p$ corresponds to the extrapolation of the linear behavior. For completeness, the dotted black line corresponds to the value of $\mathrm{SNR_{Qdyne}} \approx 88$ obtained for the fixed value of $T_2^* = 0.5$ s.

For single NV detection ($K = 1$) in Figures 4c and d, the total measurement time extended to $t_{tot} \approx 39$ h, the qualitative behavior is similar. For the achieved polarization $p = 5\cdot 10^{-5}$ in our experiments, the simulations show an $\mathrm{SNR_{CASR}} \approx 4$, already allowing for single NV detection of nanoscale samples. Note that this value of $\mathrm{SNR_{CASR}}$ lies to the left of the dotted line, which indicates that the achieved SNR is slightly lower than $\mathrm{SNR}_{\mathrm{Qdyne}}$. It is worth noting that higher polarization values can be reached using other hyperpolarization methods, such as parahydrogen-induced polarization (PHIP)[14] or dissolution DNP, that have been proven to lead to much higher sample polarizations[48], on the order of $p \sim 10^{-2} - 10^{-1}$. In that case, single-NV nanoscale sensing would be possible within experimentally feasible times, without having to worry about the resolution limitations caused by diffusion.

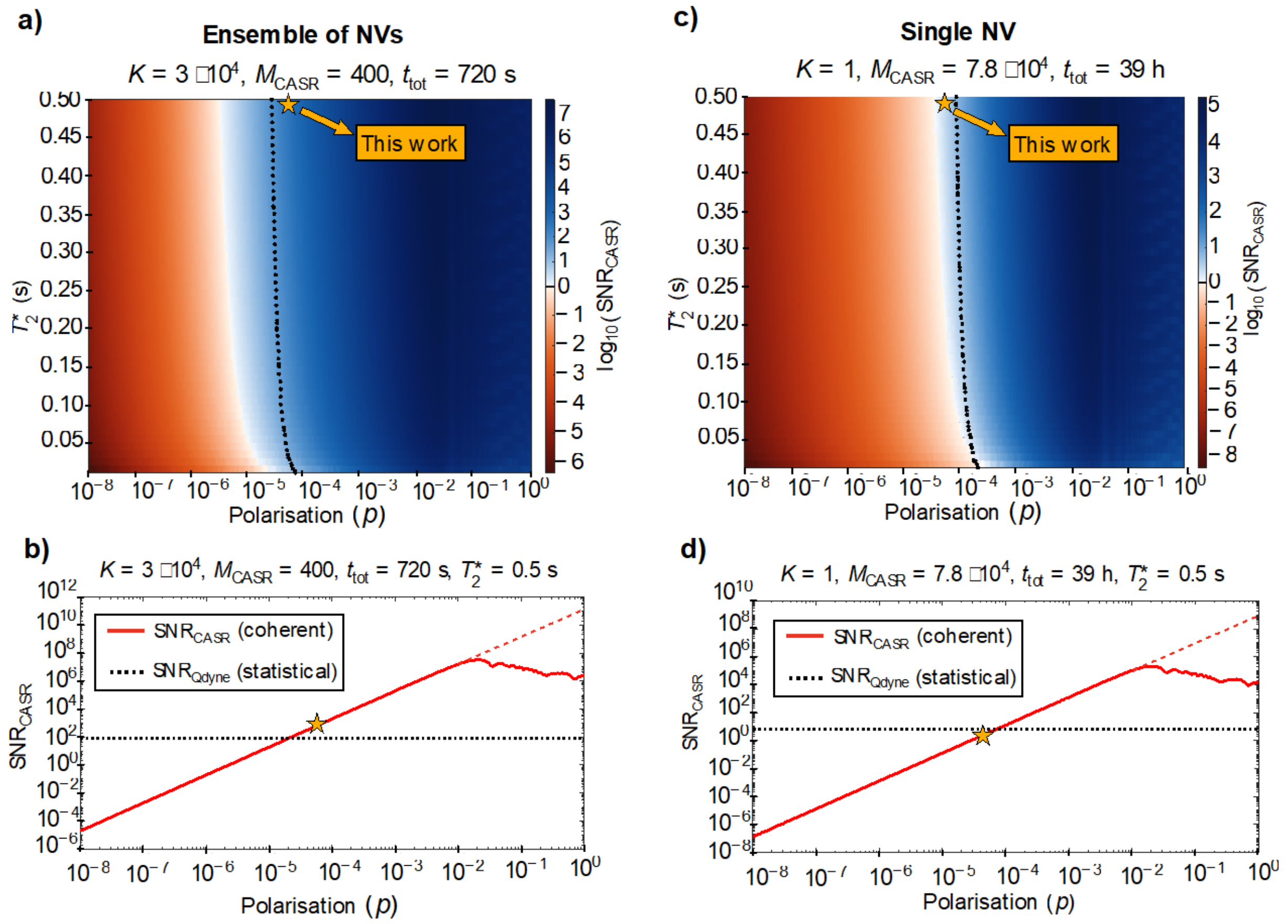

**Figure 4. Comparison of the SNRs of the CASR (coherent detection) and Qdyne (noise detection) protocols.** a) Results obtained for a NV ensemble with $K = 3 \cdot 10^4$, $\sigma_\xi \approx 6$ and $t_{tot} = 720\ s$ (that is, $M_{CASR} = 400$)**.** The plot depicts $\log_{10}(\mathrm{SNR}_{\mathrm{CASR}})$ as a function of the mean polarization $p$ and the dephasing time of the sample $T_2^*$. Warm colors indicate $\mathrm{SNR}_{\mathrm{CASR}} < 1$, whereas cold colors $\mathrm{SNR}_{\mathrm{CASR}} > 1$. The black dotted line indicates the points where $\mathrm{SNR}_{\mathrm{CASR}} = \mathrm{SNR}_{\mathrm{Qdyne}}$. The results of our experiment are marked with a yellow star. **b)** The solid red line corresponds to $\mathrm{SNR}_{\mathrm{CASR}}$ as a function of the sample polarization $p$ for a fixed value of $T_2^* = 0.5$ s. At high polarization, the small-angle approximation fails, causing a deviation from linear behavior. The red dashed line shows the extrapolation from such behavior. The black dotted line is the value for an $\mathrm{SNR}_{\mathrm{Qdyne}} \approx 88$ obtained for the fixed value of $T_2^* = 0.5$ s. **c)** and **d)** depict the results obtained for a single sensor $K = 1$, $\sigma_\xi \approx 6$ and $t_{tot} = 39$ h (that is $M_{CASR} = 7.8 \cdot 10^4$) to compensate for the lack of sensors.

**Discussion and Conclusion.** In summary, we have both simulated and experimentally demonstrated that coherent NMR signals can be detected from nanoscale liquid sample volumes, which are typically dominated by statistical polarization. This approach overcomes the challenge of diffusional line broadening[40–42], enabling high-resolution NMR experiments at the nanoscale without the necessity of sample confinement. Finally, we provided a comparative study regarding the efficiency of CASR (coherent detection) and Qdyne (noise detection) as a function of the sample polarization and $T_2^*$ showing that, under our experimental conditions, CASR is already superior.

The nanoscale detection of coherent NMR signals presents an alternative to spin noise detection, offering complementary advantages and limitations (details in SM 5). Statistical polarization detection provides large NMR signals for NVs close to the surface even at low magnetic fields and allows rapid data acquisition by eliminating the need for longitudinal relaxation time ($T_1$) recovery. However, this approach relies on variance (noise) detection, which has inherently low sensitivity, offsetting the benefit of large signal amplitudes. We hypothesize that spin noise detection will be particularly useful for probing solid-state materials[49], which are typically characterized by long $T_1$ relaxation times and high spin densities. In contrast, the detection of coherent signals at the nanoscale offers significantly higher sensitivity (slope detection) but comes at the cost of a weak NMR signal from thermally polarized samples. Hyperpolarization techniques[13,14,19,48] can overcome this limitation (at the expense of technical complexity) and, in combination with nanoscale NV detection, they may reach unprecedented spin sensitivities, in particular for liquid samples. This combination could enable localized[50–53] nanoscale sensing with high spectral resolution with applications in surface or life sciences, potentially reaching single-molecule sensitivity.

## Acknowledgments

**Funding.** This study was supported by the European Union's Horizon Europe –The EU Research and Innovation Programme under grant agreement No 101135742 ("Quench"), the European Research Council (ERC) under the European Union's Horizon 2020 research and innovation program (grant agreement No 948049) and by the Deutsche Forschungsgemeinschaft (DFG, German Research Foundation)—412351169 within the Emmy Noether program. D.B.B. acknowledges support by the DFG under Germany's Excellence Strategy—EXC 2089/1—390776260 and the EXC-2111 390814868. J.C. acknowledges the Agencia Estatal de Investigación via the Modelizado, Optimización, y Esquemas de Magnetometria en Centros de Color project PID2024-161371NB-C22, and the Basque Government grant IT1470-22. P.A.B acknowledges support from yUPV/EHU Ph.D. Grant PIF 23/275.

**Author Contributions.** D.B.B. conceived and supervised the study. N.R.v.G. performed the experiments and analyzed the data. K.D.B. wrote the code and performed the simulations of the spin dynamics. P. A. B. performed the sensitivity simulations. N.R.v.G., K.D.B., P. A. B., J.C., and D.B.B. discussed the data and wrote the manuscript with inputs from all authors.